\documentclass[journal]{vgtc}                     %
\onlineid{1333}

\vgtccategory{Research}

\title{\sys: Expressive Chart Refinement \\ Preserving Data-Binding Integrity}

\author{%
  Yumin Song, Seokhyeon Park, Soohyun Lee, Aeri Cho, Hyeon Jeon, John Joon Young Chung, and Jinwook Seo
}

\authorfooter{
  \item Yumin Song, Seokhyeon Park, Soohyun Lee, Aeri Cho, Hyeon Jeon and Jinwook Seo are with Seoul National University. E-mail: \{ymsong, shpark, shlee, archo, hj\}@hcil.snu.ac.kr, jseo@snu.ac.kr.
  \item John Joon Young Chung is with Midjourney. E-mail: jchung@midjourney.com
  \item Jinwook Seo is the corresponding author.
}

\abstract{
Creating static visualizations for presentations and publications requires granular refinements of visual details, even for simple charts.
Existing data-driven visualization tools offer limited interactive control for such refinements, forcing users to export charts to external graphic editors and breaking the critical link between data and visual representation.
To address this gap, we propose an extended InfoVis Reference Model to account for post-render design refinement.
A formative study with 18 visualization practitioners and a follow-up survey of 35 respondents confirmed that this stage is pervasive yet unsupported in current practice.
Based on these findings, we present \sys, a visualization authoring system that enables expressive visual customization while preserving data-binding integrity.
\sys supports element-level direct selection and scope expansion, allowing users to define a data-aware scope ranging from a single mark to a data-driven category with a simple selection.
For modifications beyond predefined controls, \sys{} blends natural language input with dynamically generated GUI widgets, where deictic interaction lets users reference elements simply by clicking them, keeping even open-ended edits bound to the data.
To support rigorous exploration and comparison of design alternatives, \sys implements a provenance history that enables users to capture diverse design iterations while ensuring data-visual integrity.
A user study with 12 participants verified that \sys effectively supports expressive, granular refinement without sacrificing data binding, significantly reducing repetitive manual processes in an integrated environment.

}


\teaser{
  \centering
  \includegraphics[alt={A two-part comparison diagram. The left half, labeled Current Visualization Tools, shows four problem panels. Refinement Severs Data Binding: a data table with a newly added row, joined by a broken-link icon to a bar chart annotated "2 hours in illustrator", with a user complaint about redoing every polish when data changes. Visualization Tools Lack Element-Level Control: a grey bar chart and a grey scatterplot where a cursor can select only one mark at a time, with a complaint about wanting to recolor a single bar and a single point. Predefined Controls Cannot Cover Refinement Needs: two application menu bars being searched with no matching option found. Iterative Refinement Lacks History Management: four separate chart files with version-suffixed names, and a question about which version had the right colors. The right half, labeled TailVis, shows six technique cards arranged in a grid: direct selection of one bar, scope expansion from that bar to all bars sharing a data field, deictic interaction where a clicked mark becomes a numbered token in an agent prompt, provenance history as a branching tree of chart thumbnails, attribute-aware widgets surfacing rect and axis property panels, and natural language input generating an on-demand widget with color and legend controls.},
 width=\linewidth, alt={A view of a city with buildings peeking out of the clouds.}]{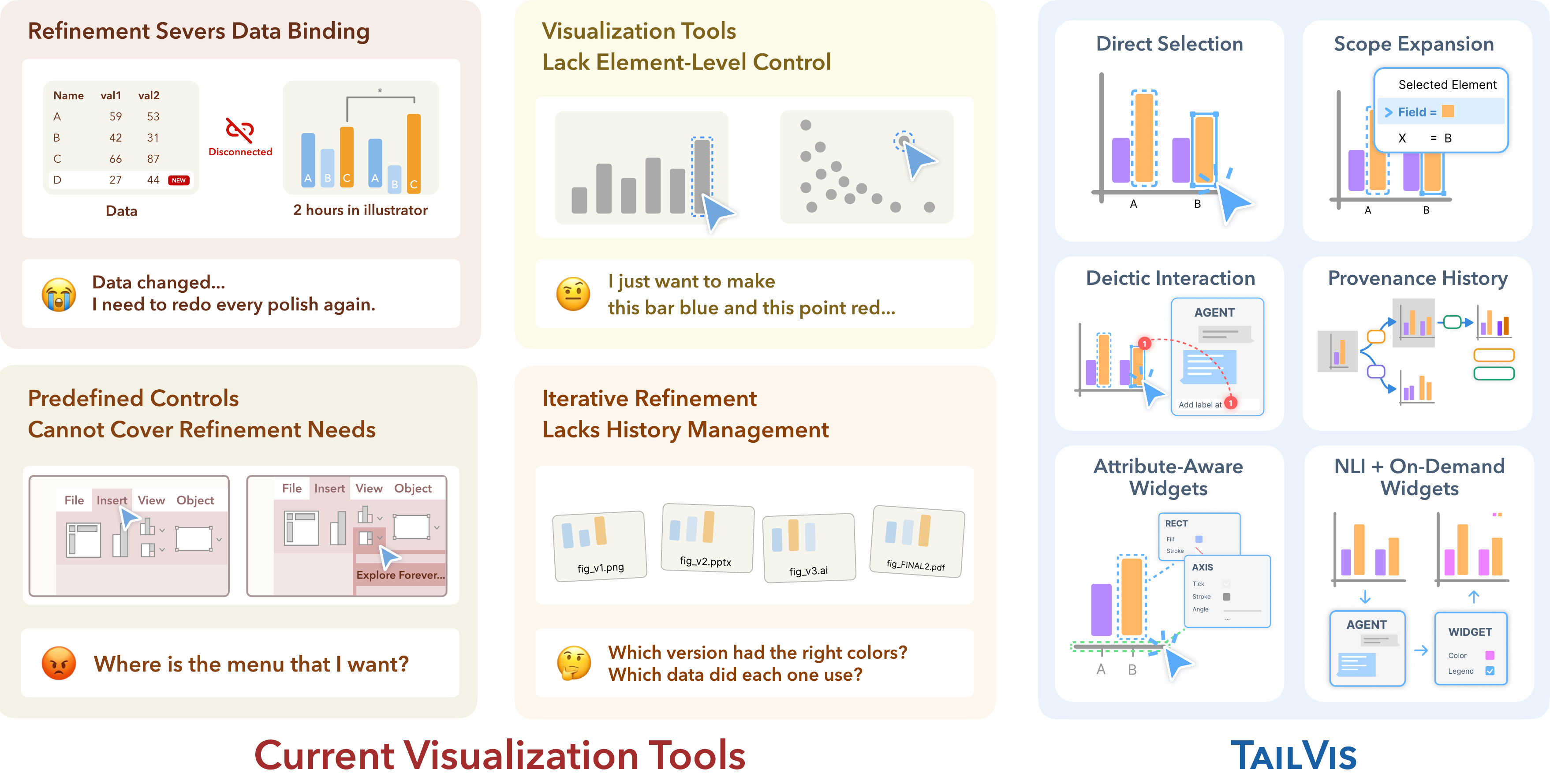}
  \caption{%
  	\sys{} enables data-bound expressive chart refinement by addressing four limitations of current practice (left): refinement that severs data binding, lack of element-level control, predefined menus that cannot cover diverse refinement needs, and absence of history management. \sys (right) provides direct selection with scope expansion, deictic interaction, attribute-aware and on-demand widgets, and provenance-integrated history, all while preserving data binding integrity.
  }
  \label{fig:teaser}
}

\graphicspath{{figs/}{figures/}{pictures/}{images/}{./}} %

\usepackage{tabu}                      %
\usepackage{booktabs}                  %
\usepackage{lipsum}                    %
\usepackage{mwe}                       %
\usepackage{ccicons}                   %
\usepackage{multirow}
\usepackage{multicol}

\usepackage{mathptmx}                  %

\usepackage{soul}
\usepackage{xcolor}

\newif\ifshowmarkup

\ifshowmarkup
  \newcommand{\del}[1]{\textcolor{red}{\st{#1}}}
  
\else
  \newcommand{\del}[1]{}
  
\fi

\newcommand{\sys}{\textsc{TailVis}\xspace}
\renewcommand{\paragraph}[1]{\vspace{1.5pt}\noindent\textbf{#1.}}

\begin{document}

\firstsection{Introduction}

\maketitle

When creating static visualizations for presentations and publications, people aim to make them visually compelling as well as informative.
People further refine the chart's visual details, such as layout, color, typography, and annotations, after initially exploring and encoding the data.
Traditionally, the visualization process has been described by the InfoVis Reference Model~\cite{card1999readings} from raw data through to the rendered view. 
However, in practice, this model does not account for the design refinement that continues beyond the initial view rendering.
Existing tools do not adequately support this refinement stage, creating a gap between data-driven generation and final visual polishing~\cite{walny2019datachanges, bigelow2014reflections, mei2018designspace, wang2023nliva}.
This disconnection in the pipeline breaks the data binding between visual elements and the underlying data, ultimately undermining the integrity and reproducibility of the entire visualization process.

Existing visualization tools fail to bridge this gap as they primarily focus on initial data transformation and mapping.
Popular software (e.g., Tableau~\cite{tableau}, Power BI~\cite{powerbi}, or Excel~\cite{msexcel}) offers basic style controls but limited support for granular adjustments.
Tasks such as freely positioning labels or independently styling individual elements remain difficult through their built-in functions.
Dedicated commercial tools (e.g., Datawrapper~\cite{datawrapper} or RAWGraphs~\cite{rawgraphs}) provide extensive refinement options; however, these remain limited to predefined controls,
leaving open-ended adjustments out of reach.
While programming libraries (e.g., D3~\cite{bostock2011d3}, ggplot~\cite{ginestet2011ggplot2}, or Matplotlib~\cite{matplotlib}) can achieve such customizations, they require significant expertise and lack interactive feedback, making iterative design refinement tedious even for skilled developers.
Consequently, many users resort to external graphic editors (e.g., Figma~\cite{figma}, Adobe Illustrator~\cite{illustrator}, or PowerPoint~\cite{powerpoint}) which allow them to freely select and manipulate each element with precision.
However, this process inevitably detaches the visualization from its underlying data~\cite{bigelow2014reflections, walny2019datachanges, bigelow2016hanpuku}.
This severance forces users to restyle elements or redo changes whenever data updates, losing data-binding as they switch between disconnected tools.

Several studies have provided ways to connect data to graphical elements~\cite{liu2018dataillustrator, xiadataink} or a glyph~\cite{ren2018charticulator}.
However, these systems focus on constructing expressive visual encodings rather than refining them, providing limited control over granular, post-render adjustments.
Moreover, many users produce common chart types for reports and presentations that do not necessarily require elaborate encoding specifications~\cite{vuillemot2017structuring}.
Bigelow et al.~\cite{bigelow2016hanpuku} attempted to bridge this gap by enabling round-trip editing between D3~\cite{bostock2011d3} and Adobe Illustrator~\cite{illustrator}, allowing users to perform detailed modifications in a graphic editor and then merge changes back.
However, their merge strategy reconstructs data, which does not stay fully bound to the graphics throughout the editing process.

To ground this fundamental issue, we propose an extended InfoVis Reference Model~\cite{card1999readings} to account for the design refinement stage that follows the rendered view.
This extension illustrates what existing visualization research has overlooked: people frequently operate beyond the rendered view, yet this stage remains disconnected from the data-bound pipeline, causing integrity and efficiency issues.

To validate our framing in real practice, we interviewed 18 researchers who regularly produce visualizations to understand their workflows, tool choices, and challenges during the design refinement, and then surveyed 35 participants with data-visualization experience to assess how broadly these issues appear in practice.
Across both studies, we found that although participants follow diverse workflows, they share a common frustration: limited refinement support in visualization tools forces repeated tool-switching, disconnecting modifications from the underlying data.
This confirms the gap highlighted by our extended InfoVis Reference Model, demonstrating that the design refinement process from a rendered view to presentation remains disconnected in current practice.
These findings informed four design goals for a data-bound refinement environment, centered on direct selection with scope control, on-demand modification, unambiguous referencing for open-ended edits, and traceable history for iterative exploration.

Guided by these design goals, we present \sys, a chart refinement system preserving data binding integrity.
Our formative study revealed that the most fundamental reason users resort to graphic editors is the ability to freely select and manipulate each element with precision, a capability absent from existing visualization tools.
\sys brings this capability into a data-bound environment, giving users granular, element-level control to directly select and manipulate chart elements such as marks, labels, axes, and annotations without breaking their data binding.

\sys structures its interaction design around four aspects.
We specify the modification target, supporting element-level direct selection and data-aware scope expansion, which propagates to its corresponding data field.
For applying modifications, \sys offers attribute-aware property widgets and natural language input with on-demand widget generation.
Deictic interaction bridges these two by letting users click elements to directly reference them in natural language commands.
Finally, a provenance history tracks every modification alongside its data state, enabling users to branch, compare, and explore design spaces.
A user study of 12 participants verifies the effectiveness of \sys compared to their original disjointed pipelines in practice.

In summary, our contributions are fourfold: (1) An extended InfoVis Reference Model that illustrates post-render visual refinement, identifying the requirements for maintaining data binding beyond the rendered view.
(2) A formative study comprising interviews with 18 practitioners and a follow-up survey with 35 participants, revealing the prevalence of data disconnection in current refinement practices and characterizing the specific barriers introduced by fragmented workflows.
(3) The design and implementation of \sys{}\footnote{The source code and live demo are available at \url{https://github.com/tailvis/TailVis}.}, a data-bound chart refinement system that introduces deictic interaction and scope expansion to support expressive, element-level modifications while preserving data-binding integrity.
(4) A user study with 12 participants that demonstrates how each interaction technique contributes to effective refinement while preserving data-binding integrity.

\section{Related Work}

\subsection{Visualization Authoring and Post-Render Refinement}

A wide range of visualization authoring tools have been proposed to help users construct charts without programming~\cite{satyanarayan2014lyra, ren2014ivisdesigner, mendez2016ivolver, ren2018charticulator, xiadataink, wang2023data, wongsuphasawat2015voyager}.
These tools offer expressive means to specify visual encodings and chart layouts, yet their interaction models are primarily designed for constructing initial visualizations rather than refining them after the data has been mapped.
For instance, Charticulator~\cite{ren2018charticulator} supports flexible glyph-based layouts through a constraint system, but modifications are mediated through templates and layout constraints rather than direct control over individual elements.
DataInk~\cite{xiadataink} enables freeform sketching and data binding through pen and touch interaction, supporting creative authoring at the data-point level.
However, these systems target the construction of novel visual representations rather than the post-render adjustment of existing charts.

Other tools address specific aspects of visual customization.
ChartAccent~\cite{chartaccent} supports annotation for data-driven storytelling, and Dupo~\cite{kim2023dupo} enables responsive visualization design through mixed-initiative recommendations.
Mystique~\cite{chen2023mystique} deconstructs existing SVG charts for layout reuse.
Beyond academic systems, commercial tools such as Datawrapper~\cite{datawrapper}, Flourish~\cite{flourish}, or RAWGraphs~\cite{rawgraphs} provide rich template galleries and styling options for producing polished charts.
However, refinement in these tools is mediated through predefined menus or templates, where users must work within their own pipeline and search for the relevant option for each property.

While each of these systems addresses a distinct customization need, none provides the general-purpose, element-level visual control that users commonly perform when preparing charts for presentation, such as freely repositioning labels, independently styling individual marks, or adjusting spacing and typography.
Consequently, despite visualization authoring tools continuing to expand what users can construct, a gap persists at the stage where users need to polish an already-rendered chart for its final output.

\subsection{Bridging Data Tools and Graphic Editors}

Studies of visualization practice have documented that users frequently export charts to graphic editors such as Adobe Illustrator~\cite{illustrator} or Figma~\cite{figma} when built-in customization falls short~\cite{bigelow2014reflections, walny2019datachanges}.
Graphic editors offer the element-level control that visualization tools lack, but this workflow detaches charts from their underlying data.
Walny et al.~\cite{walny2019datachanges} observed that once a chart is moved into a graphic editor, subsequent data changes force users to redo their modifications manually, leading to tedious and error-prone iteration.
Bigelow et al.~\cite{bigelow2014reflections} similarly reported that designers repeatedly alternate between data tools and graphic editors, with each round-trip risking inconsistency between the visualization and its data.

Several systems have attempted to bridge this disconnect.
Hanpuku~\cite{bigelow2016hanpuku} enables round-trip editing between D3~\cite{bostock2011d3} and Illustrator by inferring correspondences between modified SVG elements and the original data.
However, its binding is reconstructed after editing rather than maintained continuously, leaving room for misalignment when edits are complex and overhead that recurs across iterations.
Data Illustrator~\cite{liu2018dataillustrator} takes the opposite approach, starting from graphic primitives and introducing data binding later through a ``lazy binding'' paradigm.
This enables a designer-friendly, graphics-first workflow, but requires users to learn an unfamiliar authoring framework that differs from both conventional chart tools and graphic editors.
These approaches narrow the gap, yet neither provides a continuous data-bound environment where users can refine an existing chart with the same direct control they expect from a graphic editor.

\subsection{Blended Interaction for Visualization Editing}
Blending multiple interaction modalities in visualization systems has been actively explored~\cite{lee2015sketchinsight, chao2010napkinvis}.
Critically, natural language interfaces (NLI) have been widely adopted to lower the barrier of specifying visualization edits~\cite{shen2022towardsnli, sun2010articulate, srinivasan2021collecting, Yu2020flowsense, wang2025dataformulator}.
However, NLI-only approaches face well-known challenges: ambiguous references, lack of immediate visual feedback, and difficulty specifying spatial or element-specific operations through text alone~\cite{tory2019mean}.
To address these limitations, recent work has combined natural language with complementary modalities~\cite{srinivasan2020inchorus,  Vaithilingam2024DynaVis, l2024blended, wang2023nliva}.
DynaVis~\cite{Vaithilingam2024DynaVis} blends NLI with dynamically synthesized GUI widgets, allowing users to issue a natural language command and then fine-tune the result through a generated widget.
InChorus~\cite{srinivasan2020inchorus} combined pen, touch, and speech interactions for complementary visualization exploration on tablet devices. 
More recently, L'Yi et al.~\cite{l2024blended} formalized the design space of blended interfaces for visualization authoring, composing templates, shelf configuration, natural language, and code editing with bidirectional linking, and showed that users naturally switch among modalities to accomplish authoring tasks.
However, these blended approaches primarily target the construction or exploration stage of visualization, focusing on learnability and expressivity of initial authoring.
They do not address the efficiency and control required for post-render refinement, where users need to iteratively adjust visual details of an already-rendered chart while preserving its data binding.
In this work, we leverage blended interaction for expressive chart refinement, informed by the workflows and pain points observed in real-world visualization practice.

\section{Formative Study}
To understand how people create and refine visualizations in real-world practice,
we conducted a formative study consisting of semi-structured interviews followed by a confirmatory survey.
The study surfaced recurring challenges around data disconnection, granularity, expressiveness, and design traceability, which we translated into four design goals.
Based on the formative study results, we validate that our extended InfoVis Reference Model accounts for the design refinement process.

\subsection{Method}
We conducted a semi-structured interview study to investigate the real challenges during the visualization refinement process.
We explored their overall authoring workflows, including the tools they chose, and how they use specific functions or interface commands in those tools.
We also aimed to identify the challenges during their authoring process for the final presentation.

\subsubsection{Interviews}
We recruited 18 participants through social media advertising and snowball sampling~\cite{goodman61ams}.
They span diverse research domains, from information visualization and HCI to engineering, natural sciences, and social sciences, all of whom regularly create visualizations as part of their work.
Experience levels ranged from novice to expert (self-reported), and detailed demographics are summarized in the supplementary material.
Each semi-structured interview lasted 45--60 minutes.

Participants were asked to submit three recent visualizations beforehand as concrete prompts for recalling their authoring process~\cite{van2024understanding}.
The interview covered participants' visualization workflows from initial encoding to final presentation, the tools and strategies used at each stage, and the challenges encountered during refinement.
Four coauthors independently coded the transcripts using thematic analysis~\cite{braun2019thematicanalysis} and reconciled codes through iterative discussion.

The interviews revealed common patterns across three areas: overall workflow structure, interaction practices for refinement tasks within each tool, and challenges arising from fragmented tool use.
Detailed interview procedure and study results are in the supplementary material.

\subsubsection{Survey}

To validate the interview findings and to identify which aspects of chart refinement pose the greatest barriers, we distributed a follow-up online survey to 35 respondents with visualization experience, recruited through the same channels.
As thematic analysis characterizes which challenges arise rather than how prevalent they are~\cite{lazar2017research, Braun01012006}, we designed the survey to quantify the prevalence of the challenges identified in the interviews across a broader pool of practitioners.
We asked respondents about how often they resort to multiple tools, what motivates the switching, which refinement tasks they find most difficult, and how they handle data changes after visual edits have been applied.
Questions used multiple-choice formats; detailed questionnaires are provided in the Appendix.
Quantitative results are reported alongside the corresponding findings in \Cref{sec:findings}.
Detailed survey results are in the Appendix.

\subsection{Findings}
\label{sec:findings}
Notably, refinement strategy consistently emerged as the dominant pain point in visualization workflows across both interviews and the survey.
16 of 18 participants in the interview and 26 of 35 respondents in the survey described substantial frustration with the refinement stage, while only a few mentioned data wrangling or encoding as their primary challenge.
In the survey, when we asked an open-ended question about the single most frustrating aspect of their overall visualization process, respondents independently described issues related to refinement tasks, such as limited customization or inefficiency of fragmented tool usage.
Only 6 respondents mentioned data processing as their primary concern.
This convergence, without any specific prompt toward refinement, supports our premise that the visualization process should be understood as extending beyond the rendered view in the InfoVis Reference Model, and that this extended stage remains disconnected in current practice.
We organize the specific barriers into four findings below.

\begin{figure}
    \centering
    \includegraphics[alt={Two horizontal bar charts of survey responses from 35 respondents. Chart A, reasons for using multiple tools: another tool is easier or faster, 61 percent; polish for presentation or report, 45 percent; finer visual style control, 42 percent; add annotations or callouts, 27 percent; cannot filter or compute in tool, 9 percent; lacks data preparation capabilities, 9 percent; unsupported chart type or encoding, 6 percent. Chart B, how respondents handle data changes: manually update changed parts, 43 percent; redo from scratch, 26 percent; automated or semi-automated, 20 percent; no need to update, 6 percent; not applicable, 6 percent. Reasons related to visual polishing dominate in A, and manual strategies dominate in B.}, width=0.8\linewidth]{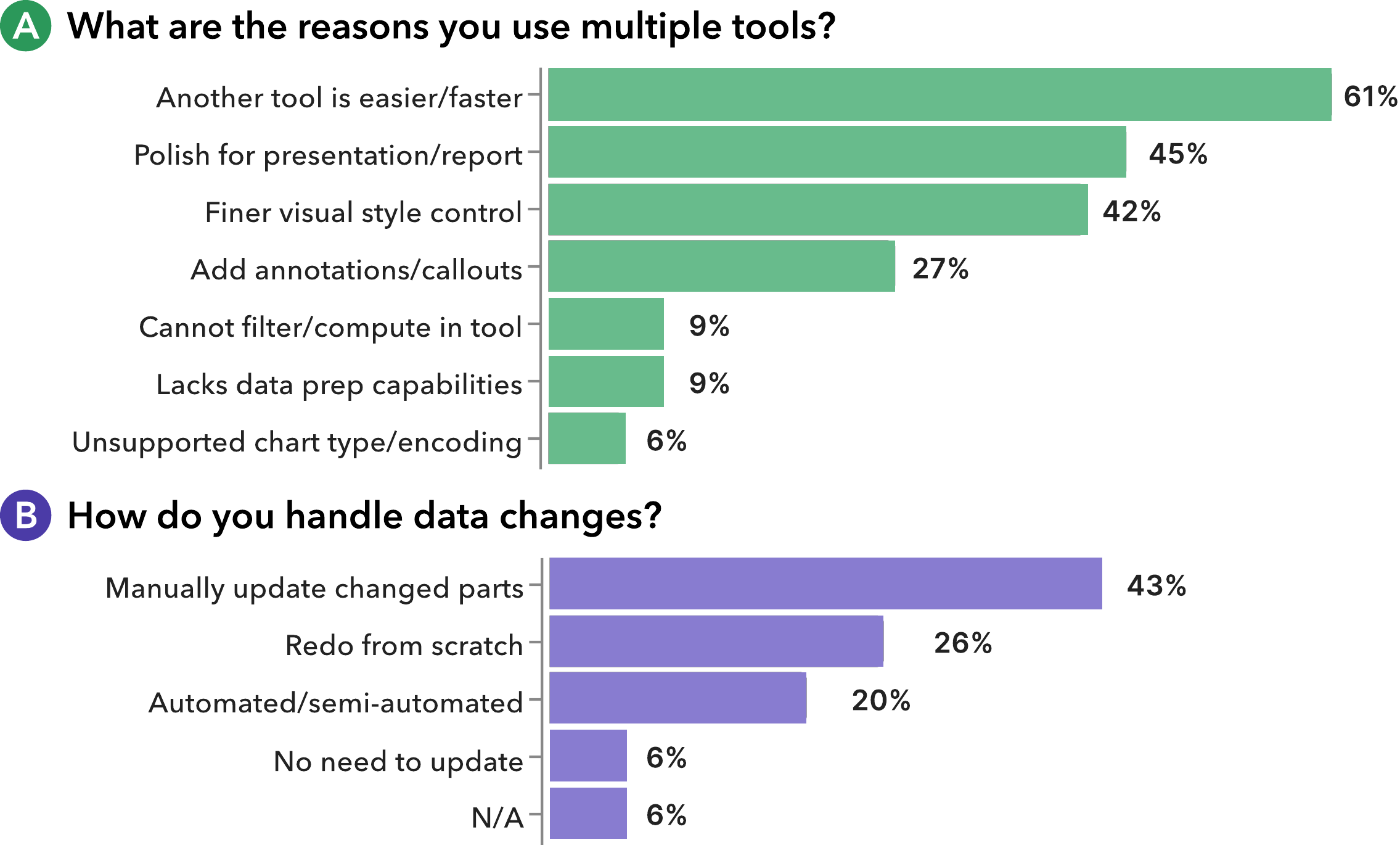}
    \caption{Survey results of two questions about using multiple tools and strategies when data changes.}
    \vspace{-2em}
    \label{fig:survey}
\end{figure}

\subsubsection{Refinement Severs Data Binding}
Across interviews, participants consistently described a pipeline in which they generate charts in code or spreadsheet tools and then move to graphic editors for design refinement.
This transition severs the connection between visual elements and the underlying data.
As P9 described, \textit{``If I switch to PowerPoint, it is hard to adjust things like placement. So if the data changes, I have to redo the latter parts, which is really annoying.''}
Our survey confirmed how widespread this pattern is: only 2 of 35 respondents reported never using multiple tools, and 69\% did so often or always.
Among the reasons for switching, the most cited were that another tool is easier or faster (61\%), polishing for presentation or report (45\%), and finer visual style control (42\%) (\Cref{fig:survey} (a)).
The cost surfaces when data changes: 43\% of respondents manually updated the changed parts and 26\% redid the chart from scratch, while only 20\% maintained an automated or semi-automated workflow (\Cref{fig:survey} (b)).
These findings show that the current refinement pipeline structurally undermines data-binding integrity.

\subsubsection{Visualization Tools Lack Element-Level Control}
In the interview, participants consistently reported that their visualization tools lack fine-grained control over individual visual elements.
P11 explained, \textit{``Because Python offers limited flexibility, I export the file to Illustrator for further editing. Labels often look unrefined, so I adjust them in Illustrator, where the higher degree of freedom helps.''}
P13 similarly noted that although MATLAB can draw legends, the text and colors look unattractive, so they recreate them in PowerPoint.
Although presentation tools such as PowerPoint are intended for composing a narrative, participants often turned to them simply because they are familiar and provide basic graphic functions. P9 noted, \textit{``I don't really know graphic editors other than PowerPoint, and it handles edits from Excel well, so rather than spend time figuring out another tool, I just restyle in PowerPoint.''}
The survey confirmed this pattern: when asked which visual properties they most frequently modify before reaching a final version, respondents selected layout and spacing (63\%), colors (57\%), resizing and repositioning (57\%), and fonts (51\%).
These are all element-level visual adjustments rather than data transformations, yet the tools participants use for chart generation do not support them adequately, forcing users into graphic editors that sever data binding.

\subsubsection{Predefined Controls Cannot Cover Refinement Needs}
Participants reported that these difficulties stem not only from limited granularity but from the absence of appropriate controls altogether.
They cited routine refinement tasks that their current tools could not support at all, such as adjusting spacing between grouped elements, reordering legend entries to match a desired reading order, or adding statistical indicators.
P6 noted, \textit{``When using Excel, it's hard to adjust things like legends or labels, even hard to find where those options are.''}
In interviews, they described not knowing where to find the function, resorting to a different tool entirely, or spending considerable time on adjustments that should have been straightforward.
The survey echoed this difficulty: the tasks rated hardest were layout and resizing (43\%), visual style customization (40\%), and handling data changes (40\%).

Although LLMs can express such open-ended modifications through natural language in a code-based environment, the 6 interview participants who had adopted LLM-assisted workflows reported persistent difficulties.
The linear conversational process made it difficult for both sides to maintain context: the model frequently misidentified the intended target or applied changes at the wrong scope, while users themselves lost track of what they requested and how each command altered the chart.
Achieving the desired result typically required multiple correction rounds, leaving these needs still inadequately addressed.

\subsubsection{Iterative Refinement Lacks History Management}
Visualization creation is inherently iterative: participants described exploring multiple color palettes, trying different font sizes, and comparing annotation styles before committing to a final design.
However, none of the interview participants used a systematic mechanism for tracking or reverting design decisions.
The dominant strategy was duplicating chart files and comparing them side by side.
One participant explained, \textit{``I always copy each chart version because I cannot just keep undoing. I try different things, compare, and change data. But as these pile up, I lose track of which visualization was made with which settings, and it becomes very hard to manage.''}
The cost of this manual tracking was amplified by tool fragmentation: reverting to an earlier design state often required not only restoring the graphic file but also re-running scripts and re-exporting from the authoring tool.

\begin{figure}
    \centering
    \includegraphics[alt={A left-to-right pipeline of five boxes connected by arrows: Source Data, Data Table, Visual Structure, View, and Presentation. The arrows are labeled underneath with the transformation applied at each step: Data Transformations, Visual Mappings, View Transformations, and Design Refinement. The first four boxes and their labels are drawn in grey, representing the original InfoVis Reference Model, while the final Presentation box and its incoming Design Refinement arrow are drawn in blue to mark the newly added stage. A horizontal blue line at the bottom, labeled Task and connected to a figure of a person, feeds upward arrows into all four transformations, indicating that user tasks influence every stage including the new one.}, width=1\linewidth]{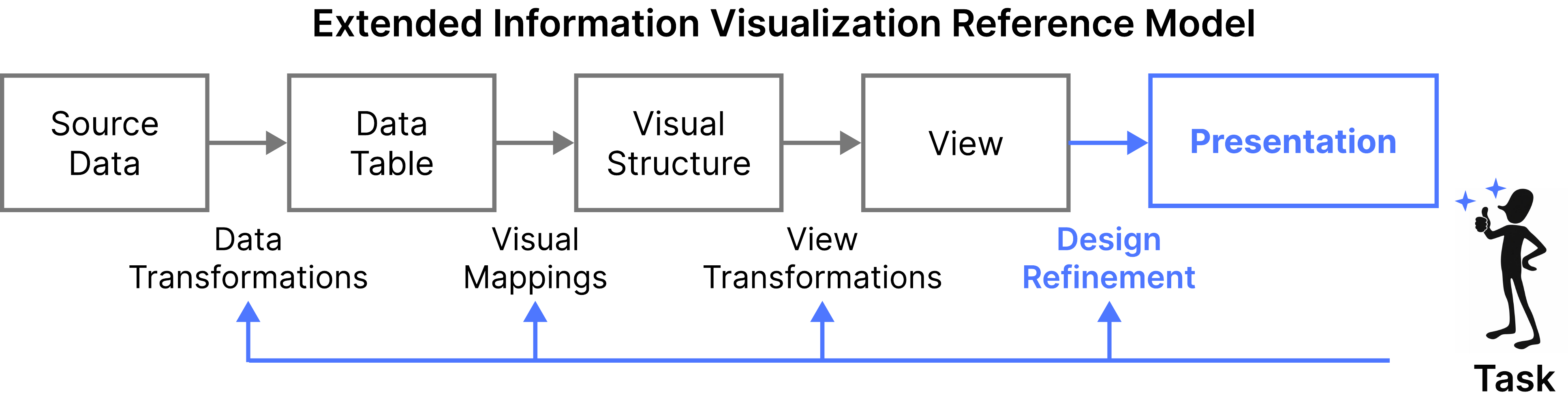}
    \caption{Extended InfoVis Reference Model. We introduce a Design Refinement stage between the rendered view and the final presentation.}
    \label{fig:referencemodel}
    \vspace{-2em}
\end{figure}
\subsection{Extending the InfoVis Reference Model}
\label{sec:model}
InfoVis Reference Model~\cite{card1999readings} describes visualization as a pipeline of successive transformations in which raw data is transformed into data tables, mapped to visual structures, and rendered as views.
This model serves as the standard conceptual framework for visualization research and tool design.
Notably, the model's final stage operates solely on what the user sees, without modifying data mappings or visual encodings.

However, as our findings show, practitioners who produce charts as presentation-ready artifacts frequently continue beyond this rendered view.
These adjustments operate on an established mapping rather than redefining how data determines visual structure, which is why the original model, whose stages each establish such a mapping, does not account for them.
We therefore extend the reference model by introducing a Design Refinement stage, specifically tailored for communicative and presentation-oriented workflows. (\Cref{fig:referencemodel}).
This stage focuses on aesthetic fine-tuning and contextualization for the final medium.
For instance, reordering legend entries to match a desired reading order rearranges the rendered output without changing how any value is encoded.
Operations that redefine the mapping itself, such as rescaling an axis, instead remain within the earlier stages.
Critically, the transition from view to presentation is where data binding often breaks in current practice: users export to graphic editors, sever the connection to data, and lose the ability to propagate data changes or verify mapping consistency.
To address this disconnect in the extended model, we derive four design goals that specify the requirements for maintaining continuous data-binding throughout the pipeline.

\subsection{Design Goals}
\label{sec:design-goals}
Based on these findings, to effectively connect this disconnection, we derived four design goals that characterize the requirements for a data-bound refinement environment.

\paragraph{DG1: Support Direct Selection and Data-Aware Scope Control}
Users resort to graphic editors primarily because visualization tools lack element-level granular control, yet this transition severs data binding.
To resolve this tension, the system should allow users to directly select and manipulate individual visual elements while preserving their connection to the underlying data.
Furthermore, when a user selects an element and issues a modification, the system should provide a mechanism to propagate that change across data-meaningful groups (e.g., all bars in the same category or all marks of the same type), rather than requiring manual repetition.
This scope control should be grounded in data attributes, not visual proximity, so that batch operations remain consistent with the data binding.

\paragraph{DG2: Provide On-demand Modification Widgets}
Participants reported that existing tools either lack fine-grained controls or make it hard to discover in complex menus.
The system should surface property controls tailored to the target element's type and attributes (e.g., radius for point marks, stroke width for lines), enabling immediate adjustment without navigating menus.
Beyond predefined controls, the system should also support open-ended modification for intents that predefined controls cannot cover, removing the need to switch to other tools.

\paragraph{DG3: Resolve Ambiguity of Intent in Open-Ended Modification}
Predefined controls cover common properties but cannot anticipate every refinement intent.
The system should also accept unconstrained user instructions to handle modifications beyond what fixed widgets offer.
However, unconstrained input introduces a new problem: users must communicate which elements to modify and how.
The system should provide means for users to communicate their modification targets clearly, reducing the gap between user intent and system interpretation.

\paragraph{DG4: Enable Design Exploration through Traceable History}
Participants lacked any systematic mechanism for tracking or reverting design decisions, resorting to file duplication and side-by-side comparison, a cost amplified by tool fragmentation.
The system should maintain a visual history that links each refinement to the underlying data state, enabling users to branch, compare, explore, and revert without losing data correspondence.
This transparency reinforces data-binding preservation: users can verify that no operation has broken data binding and can safely return to earlier states, reducing the motivation to export to external tools for version management.

\section{\sys}

\begin{figure*}
    \centering
    \includegraphics[alt={Screenshot of the TailVis web interface with seven numbered callouts. A top toolbar offers chart types such as bar, line, scatter, histogram, area, boxplot, grouped, heatmap, and pie. The center is an infinite canvas holding a line chart of Falkensee population from 1875 to 2014 with two shaded historical periods. Callout 1-A, Element-Level Direct Selection, points at a clicked legend symbol highlighted in green on the chart. Callout 1-B, Data-Aware Scope Expansion, shows a dropdown listing candidate scopes from the specific data field value through all legend symbols to the entire legend, cycled with the Tab key. Callout 2-A, Attribute-Aware Property Widgets, points to the right sidebar, which reports two selected elements and surfaces controls for position offset, fill, stroke, stroke width, and opacity, plus grouped panels for smoothing and for background and legend. Callout 2-B appears twice: once on a floating chat panel where the user types a modification in natural language, and once on a generated widget panel with dropdowns and color swatches for interpolate, line color, legend orientation, and background colors. Callout 3, Deictic Interaction, traces a green path from the clicked element on the chart into a numbered reference token inserted in the chat input. Callout 4, Provenance History, points to a small branching node graph of chart thumbnails.}, width=1\linewidth]{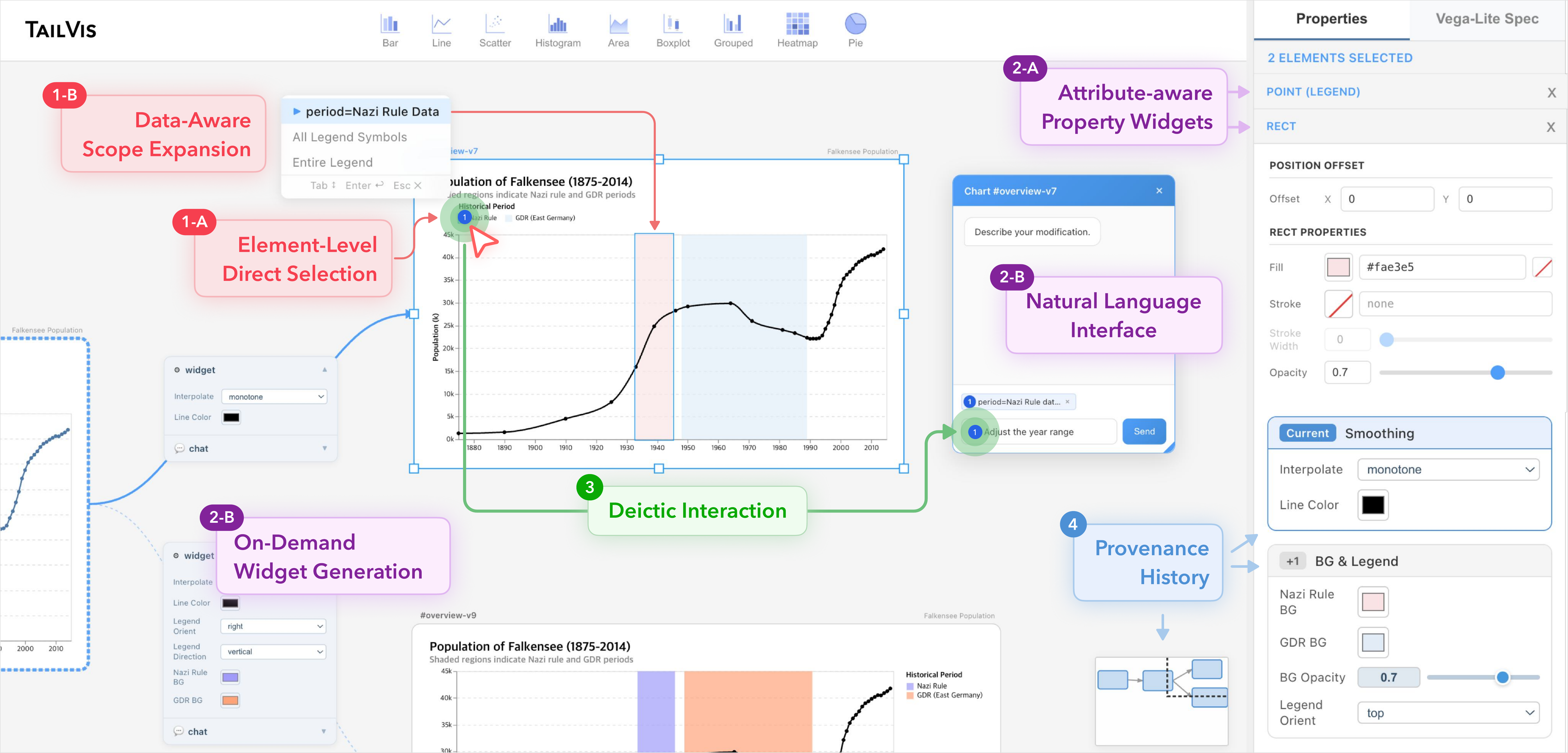}
    \caption{Overview of \sys: user interface and interaction design. (1-A) Element-level direct selection identifies the clicked element and retrieves its bound data. (1-B) Data-aware scope expansion generalizes the selection to a field-value group or all marks. (2-A) Attribute-aware property widgets surface controls tailored to the selected element. (2-B) Natural language input generates on-demand widgets for modifications beyond predefined controls. (3) Deictic interaction lets users click elements to insert reference tokens into natural language commands. (4) Provenance history records each modification as a node in a branching tree, enabling comparison and reversion.}
    \label{fig:system}
    \vspace{-1.5em}
\end{figure*}

Based on the design goals from our formative study, we present \sys{}, a system that supports expressive chart refinement while preserving data-binding integrity with element-level granular control.
In this section, we organize the interaction design around four concerns: how users specify modification targets (\Cref{sec:targeting}), express and apply modifications (\Cref{sec:modification}), bridge targeting with open-ended input (\Cref{sec:deictic-interaction}), and track and explore their design decisions (\Cref{sec:history}).

\subsection{Targeting Strategies}
\label{sec:targeting}

\subsubsection{Element-Level Direct Selection}
\label{sec:direct-selection}

\sys{} enables the selection of individual visual elements within a rendered chart.
When a user clicks any element on the canvas, the system identifies the element's type (data mark, axis label, legend entry, or annotation) and retrieves the data record associated with it (\Cref{fig:system} (1-A)).
This detection traverses the SVG structure produced by the Vega-Lite renderer, maps the clicked element to its position within the layered specification, and extracts the associated data.
The selected element serves as the target for all subsequent modification operations.
This element-level direct selection gives users granular control over each chart component, providing the same immediacy they expect from graphic editors without leaving the data-bound environment (DG1).

\subsubsection{Data-Aware Scope Expansion}
\label{sec:scope-expansion}

A single-element selection is often too narrow for refinement tasks.
Users frequently intend a modification to apply not to one mark but to all marks sharing a data attribute.
Because each element stays bound to its record,
\sys{} introduces \textit{scope expansion}, a mechanism that generalizes a point selection into a data-aware group along that binding (\Cref{fig:system} (1-B)) (DG1).

When a user selects an element, the system analyzes its data attributes and encoding structure to generate a hierarchy of candidate scopes, ranging from narrow to broad:
(1) \textit{this element}, targeting only the selected mark;
(2) \textit{by field value}, targeting all marks sharing a specific attribute (e.g., all marks where Region = ``Asia''); when a mark is bound to multiple fields, users can choose conditions (e.g., Region = ``Asia'' or Year = 2023);
and (3) \textit{all marks}, targeting every associated data mark in the layer.
Users cycle through scopes using the Tab key with immediate visual feedback.
It also applies to non-data graphical elements such as legends, axis labels, and annotations.
For example, selecting a legend entry for ``Asia'' resolves not only to the legend item itself but also to all data marks encoded with that category, bridging the graphical element and its underlying data scope.
This allows users to express targeting intent through any semantically related element.
When multiple elements are selected, the system infers shared attributes and recommends matching scopes.

Although selection is a fundamental interaction in visualization~\cite{becker1987brushing, gadhave2021predicting, ahlberg1994visual, snyder2023divi, Gadhave2024persist}, their techniques target exploration or querying rather than modification.
Query relaxation by Heer et al.~\cite{heer2008query} comes closest, generalizing a selection by relaxing data predicates, but it operates at the data level and is therefore limited to data marks, requiring manual queries once conditional mappings are introduced.
In contrast, our \textit{scope expansion} analyzes the encoding structure of a selected element to automatically generate a hierarchy of candidate scopes, extending targetable elements beyond data marks to graphical components such as axes, legends, and annotations.

\subsection{Modification Strategies}
\label{sec:modification}

\subsubsection{Attribute-Aware Property Widgets}
\label{sec:property-widgets}

Once a target is selected, the system surfaces a property widget panel tailored to the element's type and data attributes (DG2) (\Cref{fig:system} (2-A)).
Rather than presenting a fixed toolbar, \sys{} inspects the selected element and displays only the relevant controls: radius and fill for point marks, stroke width and interpolation for lines, corner radius and opacity for bars, font properties for text elements, and so on.
For structural elements such as axes, legends, and chart titles, the system surfaces the corresponding specification-level properties (e.g., axis title, label angle).
The widget set draws on common controls from graphic editors and visualization tools to minimize learning cost.

When scope expansion is active, the property widgets also reflect the scope: modifying a property at the \textit{by field value} scope inserts a conditional encoding rule, while modifying at \textit{all marks} updates the property globally.
For example, selecting an individual legend label surfaces controls for that label's mark type, while expanding the scope to the entire legend surfaces structural properties such as background, padding, and orientation.
These widgets enable adjusting common visual properties with immediate feedback, without searching through menus or composing instructions.
All changes are written into its Vega-Lite specification, preserving data binding.

\subsubsection{Natural Language with On-Demand Widget Generation}
\label{sec:nl-widgets}

Attribute-aware property widgets cover common properties but cannot anticipate every refinement intent.
For modifications beyond what fixed widgets offer, \sys{} accepts natural language instructions and generates persistent, interactive widgets from them (DG2).

When a user issues a natural language command, the LLM backend produces a modified Vega-Lite specification together with a set of widget definitions (\Cref{fig:system} (2-B)).
Each widget specifies a control type (slider, color picker, dropdown, text, or toggle), and a path pointing to the corresponding property in the specification.
Interacting with a generated widget directly updates the specification at the bound path, allowing iterative adjustment without reissuing the command.

Structuring natural language input and synthesizing dynamic widgets from it has been shown to improve the efficiency of iterative refinement~\cite{snui, Vaithilingam2024DynaVis}.
However, its one-shot generation provides no preview, and as widgets accumulate in a linear stack, their binding to properties becomes hard to track and may break when the specification changes.
We instead reframe the LLM as a chart-specific agent that proposes modifications through interactive widgets, each recorded in the provenance history (\Cref{sec:history}) so the user can revisit and adjust any prior state without reissuing the command.

\subsection{Deictic Interaction}
\label{sec:deictic-interaction}

The targeting strategies in \Cref{sec:targeting} let users precisely specify elements and groups, but this precision is not directly available when composing natural language commands.
Users must describe targets in text, which is verbose and error-prone in visually dense charts (DG3).

\sys{} bridges this gap through \textit{deictic interaction}: users click elements before or during a natural language command, inserting numbered reference tokens into the input.
The system extracts the data and visual properties bound to each referenced element and passes them as structured context to the LLM (\Cref{fig:system} (3)).

Deictic reference has been widely used in prior work to identify objects through verbal, spatial, or sketch-based input~\cite{bolt1980put, kobsa1986combining, wang2023nliva, heer2008query,han2024deixis}.
In \sys{}, we use deictic reference to bridge direct targeting and natural language modification.
When using the natural language interface, users can insert selected elements as reference tokens into the prompt, providing the LLM with the exact data records and encoding properties bound to those elements.

This enables concise, spatially grounded commands such as ``make \texttt{[1]} the same color as \texttt{[2]}'' or ``add a label to \texttt{[1]},'' replacing verbose descriptions like ``change the color of the bar representing Asia in 2023 to match the bar representing Europe in 2023.''

Deictic interaction also combines with scope expansion.
A user can reference an element, expand the scope to its categorical group, and issue a command that applies to the entire group.
This resolves two sources of ambiguity simultaneously: the deictic reference identifies \textit{which} element, while scope expansion determines \textit{how} the modification propagates.
Either way, users reach the underlying datum through the graphical element, keeping the command bound to data.

\subsection{Provenance History}
\label{sec:history}
Visualization refinement is iterative: users explore alternative styles, compare options, and revisit earlier decisions.
Our formative study found that people have no reliable way to track these changes, resorting to copying files and losing track of which version used which settings.

\sys{} records every modification as a node in a branching derivation tree (\Cref{fig:system} (4)).
Each node stores the Vega-Lite specification, the command or action that produced it, the generated widgets, and a visual thumbnail.
\sys{} separates history into two levels to avoid cluttering the tree with minor edits.
Generating a new widget through the NLI creates a new node, while fine-grained adjustments through property widgets are recorded as a history stack within the current node.
Both are trackable and revisitable: users can navigate to any node and replay or remove individual stack entries without affecting subsequent changes, without the need to overwrite history (DG4).

Branching history has proven effective for non-linear workflows, letting users revisit, compare, and recover without overwriting their work~\cite{heer2008history, Freire2006sciworkflow, Bavoil2005VisTrail}.
These models, however, have been developed primarily for analytic exploration, tracking the reasoning path through a dataset rather than the evolution of a chart's design~\cite{Gadhave2024persist, Yu2020flowsense, Yu2017VisFlow}.
\sys{} applies this model by embedding the modification action, the generated widgets, and the associated data state within each node, not just the resulting specification.
This makes the history view a traceability mechanism: users can inspect any node to verify that its modifications remain consistent with the underlying data.

Across the branches, a data update propagates to every node rather than staying local to one state.
A modification applied to a data-defined group is stored as a conditional encoding rule, so it extends automatically to new marks satisfying the condition, while a modification tied to a single element stays bound to its specific datum.
Non-data-bound elements follow their anchor: an annotation on a computed value such as a series maximum relocates as the value changes, while one tied to a specific datum keeps its position and updates only its label.
When a styled element is removed, its modification becomes inactive rather than discarded and reapplies if the element reappears.
Refinements thus remain bound to the data rather than the rendered output, so design work persists through data changes.

\subsection{Implementation}
\label{sec:implementation}

\sys{} is a web application with a React frontend and a Node.js backend.
Charts are rendered using Vega-Embed on an infinite canvas with D3-based pan and zoom.
We adopt Vega-Lite as the specification format: its declarative grammar represents charts as JSON with explicit data-to-visual mappings, ensuring that modifications operate on the specification rather than the rendered SVG.

Element detection operates over the rendered SVG tree.
When a user clicks an element, the system traverses the SVG hierarchy to classify the element type, extract the bound datum, and determine its layer position.
Scope expansion is computed client-side by analyzing the element's datum against the chart's encoding structure; each scope translates into a conditional encoding predicate.

The backend exposes API endpoints for chart creation, modification, and chat interaction, powered by Claude Opus 4.8.
Each endpoint receives the current specification with a statistical data summary and returns a modified specification with widget definitions.
Edits are validated through Vega-Lite's compiler before application; failed validations trigger automatic retries with error feedback.

\section{User Study}
\label{sec:evaluation}
 
We conducted a within-subjects user study with 12 participants, comparing \sys{} against a baseline system to evaluate whether its interaction techniques improve the chart refinement experience while preserving data-binding integrity.
The study was approved by the university IRB.

\subsection{Participants}
\label{sec:participants}
 
We recruited 12 participants (5 female, 7 male; aged 23--32, M=27.3) through advertisements in online communities of local universities and snowball sampling~\cite{goodman61ams}.
Participants came from diverse fields, including HCI, data visualization, robotics, sports science, linguistics, and business consulting.
All regularly create charts for academic or professional purposes using tools such as Excel, PowerPoint, Python (matplotlib~\cite{matplotlib}, ggplot~\cite{ginestet2011ggplot2}), Figma, Illustrator, Matlab, and R.
Self-rated chart authoring proficiency ranged from 1 (basic chart generation) to 5 (free-form design implementation) on a five-point scale, with a median of 3.

\subsection{Study Design}
\paragraph{Baseline}
\label{sec:baseline}
Our goal was to isolate the effect of \sys{}'s targeting and refinement techniques rather than to compare overall tool productivity.
Our formative study showed that practitioners follow highly diverse workflows, and that their central difficulty lies in controlling modifications rather than expressing them (\Cref{sec:findings}).
This diversity also makes any single user's existing workflow unsuitable as a controlled baseline, as it would vary in tools, feature sets, and proficiency across participants.
We selected DynaVis~\cite {Vaithilingam2024DynaVis} as a baseline, an LLM-powered editing tool whose blended interaction of natural language input and dynamically generated GUI widgets has been empirically validated for iterative chart modification.
Because \sys{} adopts the same NLI-to-widget mechanism as its modification strategy (\Cref{sec:nl-widgets}), DynaVis provides a controlled foundation that shares this interaction while isolating the techniques we contribute, without requiring tool-specific commands.
This enables fairer comparison across varying levels of tool experience and visualization literacy.
To control for differences in LLM performance and widget generation quality, both systems were configured with the same underlying language model and prompt logic.

\begin{figure}
    \vspace{-0.5em}
    \centering
    \includegraphics[alt={Four study charts arranged in two groups. The Task 1 group on the left shows two original charts each overlaid with a card listing refinement sub-tasks: chart A is a scatterplot of penguin body mass against flipper length colored by three species, and chart B is a histogram of binned measurements with control and treatment groups in blue and orange. The Task 2 group on the right shows two before-and-after pairs joined by arrows: chart C goes from a plain blue line chart of Falkensee population to a styled target with shaded regions marking historical periods, a subtitle, and a legend; chart D goes from a plain bar chart of monthly nonfarm payroll changes to a styled target with a descriptive headline, shaded presidential-term regions, a dashed divider, and reformatted axis labels.}, width=1\linewidth]{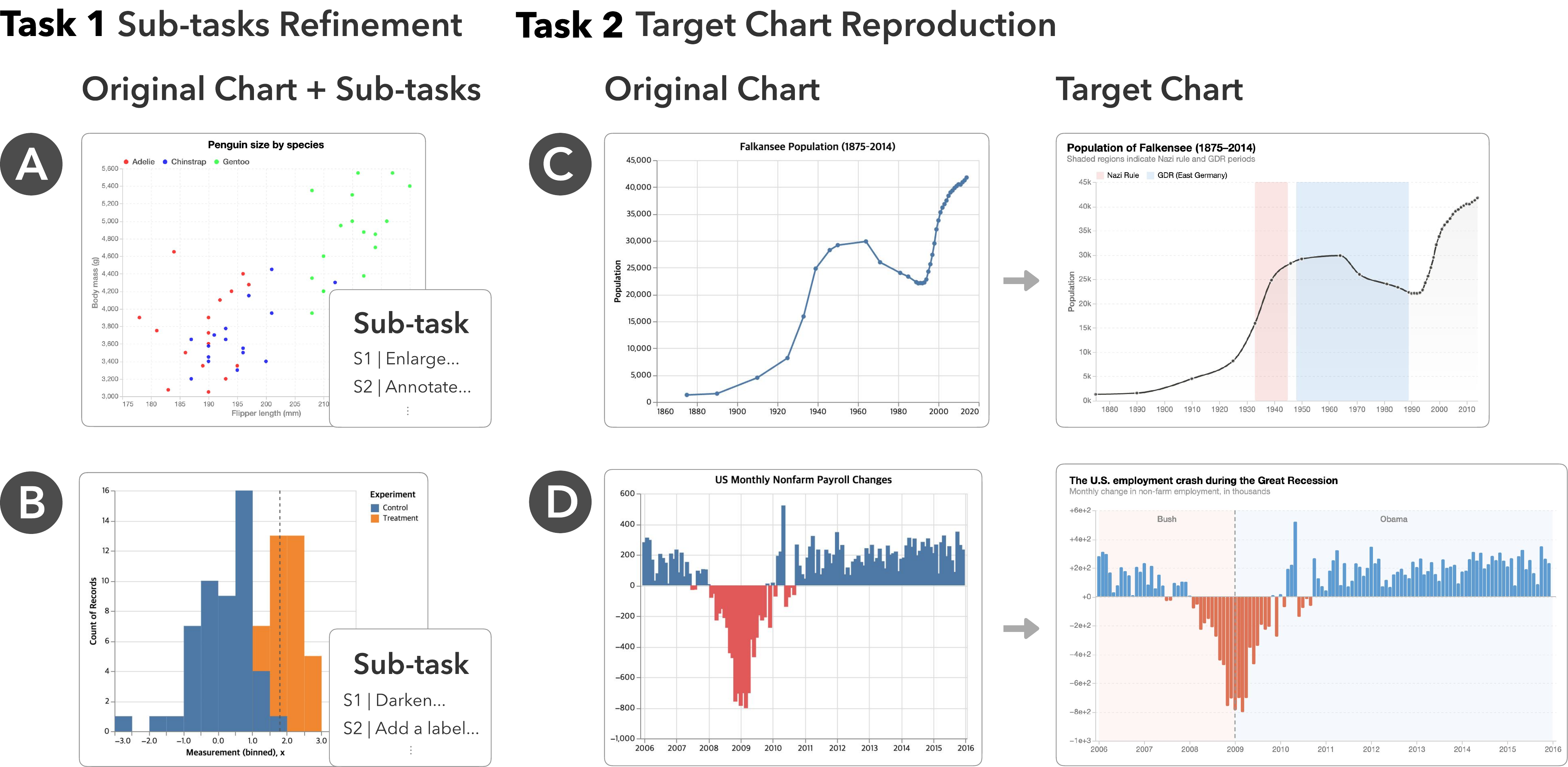}
    \caption{Summary of user study tasks. Task 1 is a multi-subtask refinement (12 min). Task 2 is a target-image reproduction (15 min). Each participant completed both tasks under the baseline and \sys{} conditions, counterbalanced using a Latin square design.}
    \label{fig:task}
    \vspace{-0.5em}
\end{figure}

\begin{table}[t]
\centering
\caption{Subtasks for Task 1. Tasks use visual/spatial descriptions to prevent direct copying into the NL interface.}
\label{tab:subtasks}
\small
\begin{tabular}{@{}c @{}c p{0.82\columnwidth}@{}}
\toprule
\textbf{Chart} & \textbf{\#} & \textbf{Sub-task} \\
\midrule
\shortstack{(a)}
 & S1 & Enlarge the most separated species; lower the other two's opacity. \\
 & S2 & Annotate the outlier in the Adelie group with a circle and label. \\
 & S3 & Try combinations of legend position, chart size, and color; pick the best. \\
 & S4 & Emphasize three upper-right points with larger size. \\
\midrule
\shortstack{(b)}
 & S1 & Darken the higher-mean group's bars; lighten the other group. \\
 & S2 & Add a label above the dashed line (``Treatment mean = 1.80''). \\
 & S3 & Try combinations of opacity, border style, and color; pick the best. \\
 & S4 & Add a dashed line and label at the Control group's mean position. \\
\bottomrule
\end{tabular}
\vspace{-2em}
\end{table}

\paragraph{Tasks}
\label{sec:tasks}
We designed two task types to cover complementary aspects of design refinement, rather than full chart authoring.
Across both tasks, we prepared four charts of different types (scatterplot, histogram, line chart, and grouped bar chart), each with different datasets (\Cref{fig:task}).

\noindent\textit{Task~1: Sub-tasks refinement.}
Informed by the task design of DynaVis, Task~1 consists of a series of 4 sub-tasks to edit the visualization.
Participants received a base chart and a list of refinement subtasks (\Cref{tab:subtasks}).
In Task~1, we intentionally phrased sub-tasks using visual or spatial descriptions rather than explicit data labels.
For example, instead of describing a task including a specific field name like ``enlarge the point GENTOO group'', we used ambiguous descriptions such as ``Enlarge the most separated species'' to prevent participants from simply copying task instructions into the natural language input.
Tasks also include a design exploration sub-task (\Cref{tab:subtasks} S3) that requires multiple combinations, reflecting iterative refinement in practice.
This task evaluates how well each system supports a sequence of targeted modifications that require identifying and selecting specific elements.

\noindent\textit{Task~2: Target chart reproduction.}
To explore interaction behavior of participants beyond the structured sub-tasks, we also designed target visualization reproduction task.
Participants were given a base chart and a reference visualization of the desired final state.
They were asked to modify the chart to match the target as closely as possible including color, legend, annotation, and size.
This task evaluates whether users can achieve a specific visual outcome through their own strategy, requiring both precise adjustments and exploration of available controls.

\paragraph{Procedure}
Each session lasted approximately 90 minutes and was conducted remotely via Zoom with screen and audio recording.
After informed consent and a demographic questionnaire, participants received a 5-minute tutorial on the first assigned system.
They then completed one Task~1 and one Task~2 with the first system (12 and 15 minutes, respectively).
After switching to the second system and receiving another tutorial, they repeated both tasks with different charts.
To mitigate learning effects, we counterbalanced both the system order and chart assignments across participants using a Latin square design.
 
After both conditions, participants completed the UEQ-S (User Experience Questionnaire, Short Version)~\cite{hinderks2017UEQ} for each system and a post-survey on chart authoring practices.
We then conducted a semi-structured interview (approximately 20 minutes) covering four themes: overall experience comparison, feature-level interaction, refinement process, and system evaluation.

\subsection{Measures and Analysis}
\label{sec:measures}
 
We collected both quantitative and qualitative data.
Quantitative measures included UEQ-S scores for pragmatic and hedonic quality, and task completion logs capturing timestamps and interaction strategies.
For qualitative analysis, four coauthors independently coded the interview transcripts using thematic analysis~\cite{braun2019thematicanalysis}.
Codes were reconciled through iterative discussion and organized around the four design goals.

\section{Results and Findings}
In this section, we report our user study findings including general usability and usefulness, task completion statistics, and user behavior.

\subsection{General Usability and Usefulness}
\label{sec:usability}

\begin{figure}
    \centering
    \includegraphics[alt={A diverging stacked bar chart of UEQ-S responses on a seven-point semantic differential scale. Eight rows pair a negative anchor on the left with a positive anchor on the right: obstructive versus supportive, complicated versus easy, inefficient versus efficient, confusing versus clear, boring versus exciting, not interesting versus interesting, conventional versus inventive, and usual versus leading edge. Each row contains two bars, one for the baseline and one for TailVis, with responses stacked around a common center. Across all eight items the TailVis bars sit further toward the positive anchor than the baseline bars, and the gap is widest for efficient, exciting, interesting, inventive, and leading edge. The complicated versus easy and confusing versus clear rows show the most overlap between conditions.}, width=0.85\linewidth]{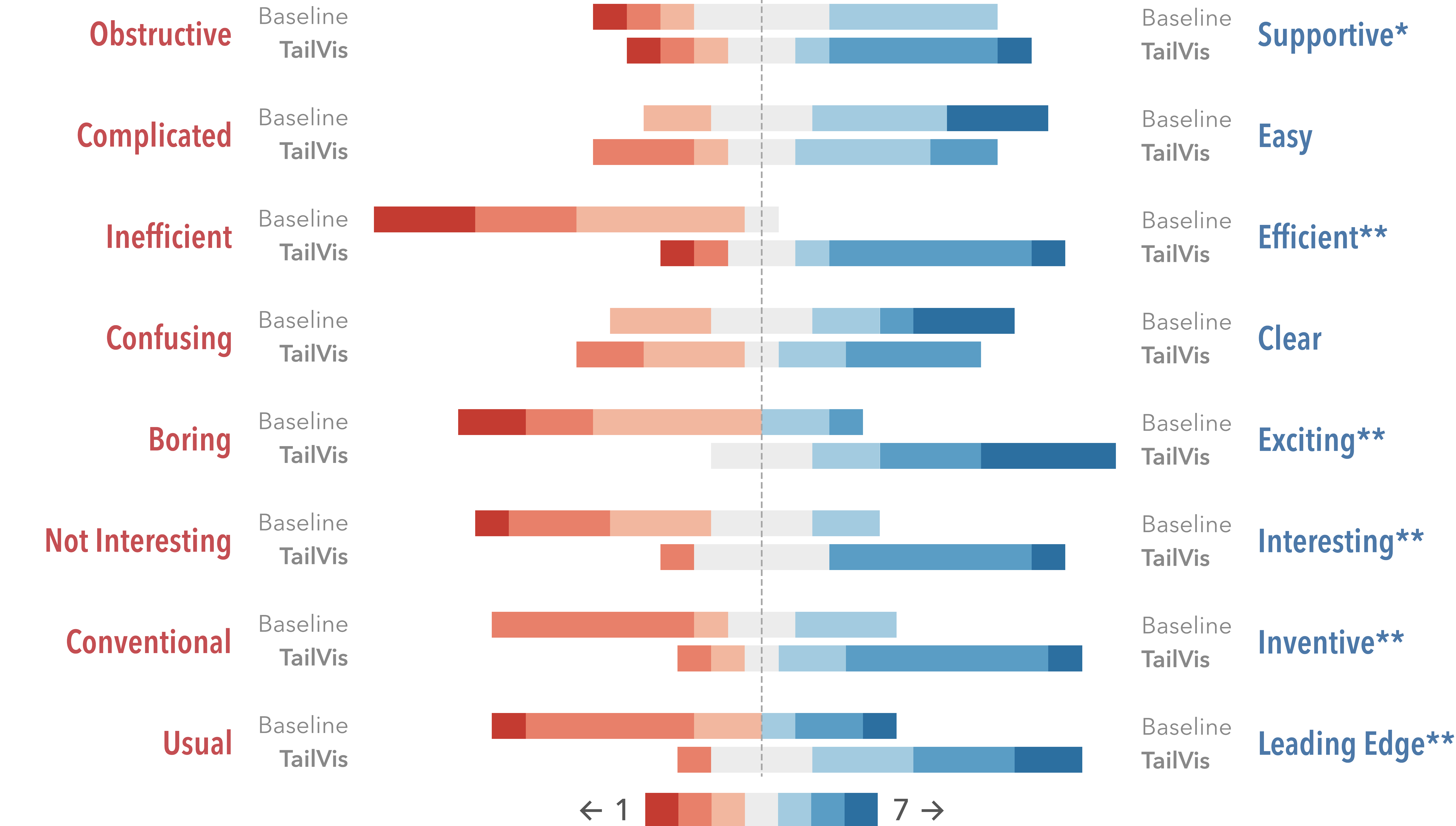}
    \caption{UEQ-S results. Significant differences on 6 of 8 items; non-significant items reflect initial learnability.}
    \label{fig:UEQ}
    \vspace{-1em}
\end{figure}

\begin{figure}
    \centering
    \includegraphics[alt={A diverging stacked bar chart of five custom usability items rated on a seven-point scale, each with a baseline bar and a TailVis bar. The items are expressiveness, control, learnability, intent recognition, and satisfaction. TailVis responses concentrate toward the high end of the scale for expressiveness, control, and intent recognition, while the baseline responses for those items spread further toward the low end. The learnability rows for the two conditions are nearly identical, and the satisfaction rows differ only slightly in favor of TailVis.}, width=0.85\linewidth]{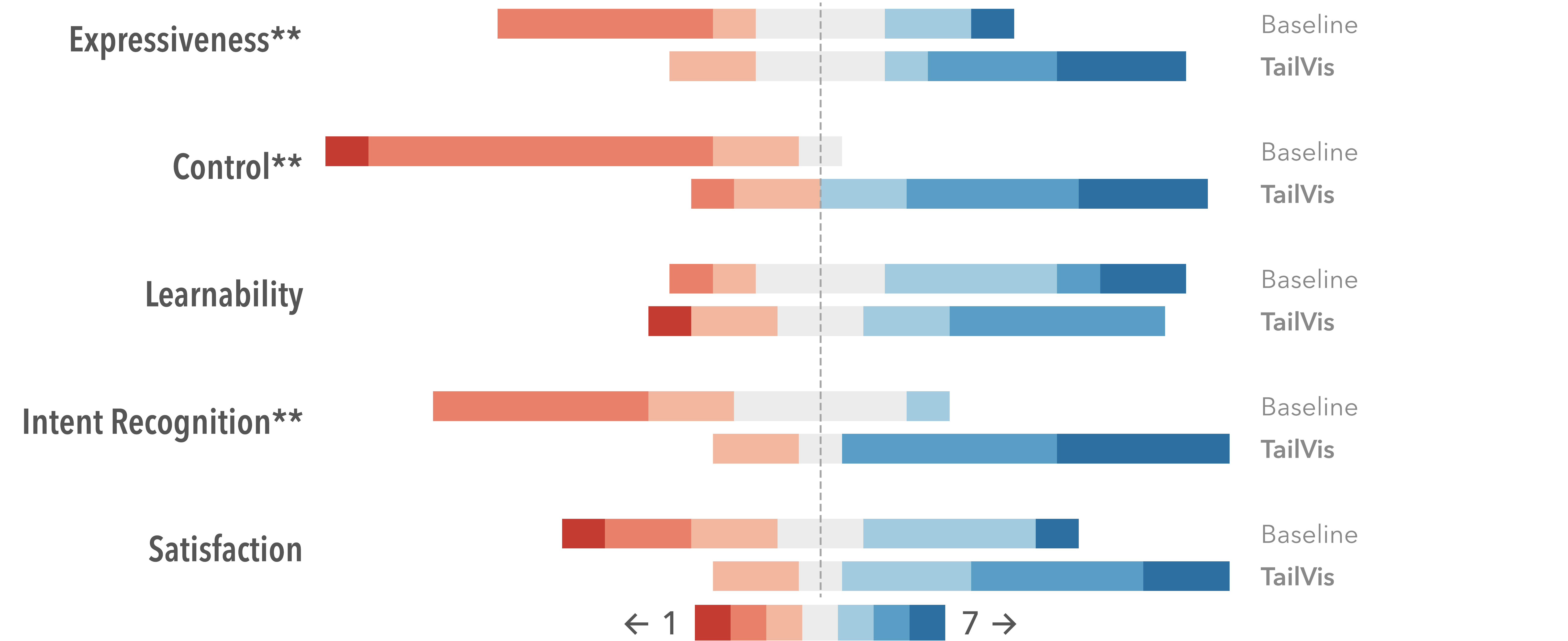}
    \caption{Custom usability results. \sys scored significantly higher on Expressiveness, Control, and Intent Recognition.}
    \label{fig:usability}
    \vspace{-2em}
\end{figure}

We compared UEQ-S (Short User Experience Questionnaire)~\cite{hinderks2017UEQ} 7-point semantic differential scores (\Cref{fig:UEQ}).
Given the paired design of our study, we employed Wilcoxon signed-rank tests~\cite{wilcoxon1945individual} for all statistical comparisons.
The results reveal significant differences ($*p<.05$, $**p<.01$) for 6 of 8 items, with particularly strong effects for efficient ($p=.0039$) and interesting ($p=.0034$).
This confirms that participants found \sys{} more effective and engaging for chart refinement than the baseline.

We also compared five usability items on a 7-point Likert scale (\Cref{fig:usability}).
\sys{} scored significantly higher for 3 of 5 items: expressiveness ($p=.0078$), control ($p=.0010$), intent recognition ($p=.0020$).
Satisfaction showed a trend favoring \sys{} but did not reach significance ($p=.094$). Learnability did not differ between conditions.
These results indicate that participants felt \sys{} provided better control and more accurately recognized their editing intent.

Across both measures, the non-significant items share a common pattern: complicated/easy and confusing/clear in UEQ-S, and learnability in the usability questionnaire, all relate to initial ease of use.
Participants attributed this to the short familiarization period with \sys{}'s broader feature set, while the baseline involves only NL input and widget interaction.
Post-interviews confirmed this was a learnability effect: participants consistently reported that the system became easier with continued use and expected the gap to diminish with longer exposure.

\subsection{Task Performance}
\textit{(Task~1) Sub-tasks achievement rate.}
The first author and a co-author independently rated whether each sub-task was achieved, resolving disagreements through discussion.
Participants using \sys{} achieved a higher proportion of sub-tasks ($\mu=0.897$, $\sigma=0.130$) than those using the baseline ($\mu=0.694$, $\sigma=0.231$), a statistically significant difference ($p=0.035$).
Example results are in the Appendix.
For individual charts, all comparisons showed consistent trends favoring \sys{}: \Cref{fig:task}~(b) showed the largest gap (DynaVis: $\mu=0.667$, $\sigma=0.249$; \sys{}: $\mu=0.933$, $\sigma=0.094$), while \Cref{fig:task}~(a) showed a smaller but directionally consistent difference (DynaVis: $\mu=0.722$, $\sigma=0.208$; \sys{}: $\mu=0.861$, $\sigma=0.150$).
Because each chart had only six participants per condition, this lacked sufficient power for reliable significance testing; we therefore report the aggregate result as the primary measure.
We did not analyze completion time, as Task~1 includes open-ended exploratory sub-tasks.

\begin{figure}
    \centering
    \includegraphics[alt={Two box plots of per-participant interaction proportions in TailVis, with individual participant points overlaid. Panel A, input modality, has three boxes: element select with a mean of 57.9 percent and the widest spread, scope confirm at 25.0 percent, and NL prompt at 17.1 percent with the narrowest spread. Panel B, modification strategy, has two boxes: property widget at 67.9 percent and dynamic widget at 32.1 percent. Direct targeting actions clearly outweigh natural language prompts, and built-in property widgets outweigh LLM-generated dynamic widgets.}, width=0.7\linewidth]{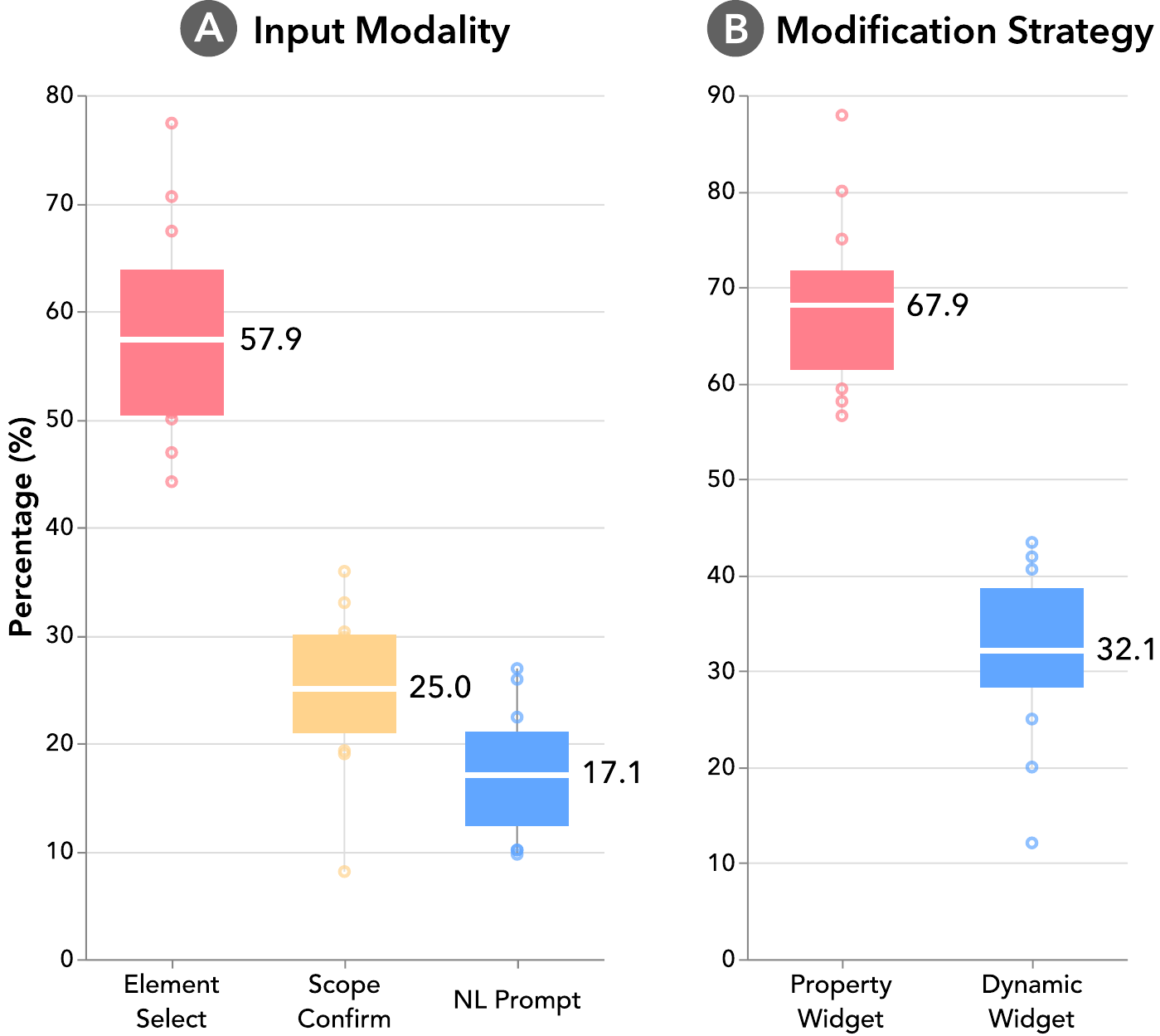}
    \caption{Proportion of interaction counts in \sys{}: (a) input modality across targeting strategies and NL prompts, (b) modification strategy across widget adjustments and property changes.}
    \label{fig:interaction}
    \vspace{-2em}
\end{figure}

\noindent\textit{(Task~2) Completion time and completion rate.} Most participants used nearly the full allotted time in both conditions (DynaVis: $\mu=14.13$\,min, $\sigma=1.39$; \sys: $\mu=13.72$\,min, $\sigma=1.42$), reflecting the open-ended nature of the task. Despite the shorter familiarization period with \sys{}'s broader feature set, participants performed significantly more modification actions in \sys{} than in the baseline ($p{=}.045$), with approximately 40\% more combined widget and property adjustments. As shown in \Cref{fig:interaction} (b), participants tended to accept LLM-generated widgets with minimal changes while actively exploring design alternatives through property widgets, suggesting deliberate exploration rather than corrective overhead. The two conditions yielded no significant difference in CLIP-based semantic similarity to target charts (DynaVis: $\mu=0.656$, $\sigma=0.017$; \sys: $\mu=0.654$, $\sigma=0.024$). Because CLIP embeddings capture high-level semantic layout rather than the fine-grained style differences that element-level refinement targets, we treat Task~2 primarily as a probe for interaction behavior (\Cref{sec:userbehavior}) rather than a measure of output fidelity.

\subsection{User behavior}
\label{sec:userbehavior}

\subsubsection{Targeting Strategies Enhance Controllability}
Participants consistently reported that the ability to select elements before modifying them was a primary factor in their sense of control.
Even when \sys{} provided only basic options on property widgets, participants perceived higher granularity simply because they could specify \textit{what} to modify before specifying \textit{how}: \textit{``being able to adjust at the right granularity stood out''} (P2), \textit{``I couldn't choose what to change in DynaVis, but I felt more comfortable in \sys''} (P12).

For scope expansion, 7 of 12 participants explicitly mentioned that it reduces unnecessary overhead; P9 and P4 noted that they used it most frequently, and P10 valued \textit{``catching everything at once rather than clicking one by one.''}
\Cref{fig:interaction} (a) shows the proportion among targeting strategies and NL prompts.
Across sessions, participants relied on targeting strategies more often than natural language prompts, with scope expansion accounting for a large portion of element selections.
We counted only targeting actions that led to a subsequent property change.
As expected, these targeting strategies also reduced reliance on natural language input.
Participants in \sys{} issued significantly fewer NLI commands ($\mu=12.0$) than those using DynaVis ($\mu=20.0$, $p=0.001$).

\subsubsection{On-demand Modification Affords Expressive Flexibility}
In \sys{}, selecting an element immediately surfaces relevant property widgets without requiring users to search for applicable controls.
This allowed participants to handle straightforward adjustments directly through widgets and to delegate unfamiliar operations to NLI, choosing the appropriate modality for each task.
P3 noted \textit{``widgets appear right away on selection, so I did simple changes manually and only used NLI for things not in the panel.''}
P5 commented \textit{``in DynaVis, changing legend color only gives one widget and adjusting position requires another prompt, but \sys{} shows all relevant options at once.''}

This combination of readily available widgets and open-ended NLI broadened the range of modifications participants could express within a single environment.
P6 mentioned \textit{``even in PowerPoint or Illustrator, figuring out whether a certain feature exists is itself a burden,''} suggesting that on-demand widget lowers the threshold for attempting expressive refinements that users might otherwise not pursue.
By offering both property widgets and dynamic widgets, \sys{} lets users quickly access common modifications through direct controls while handling open-ended tasks through natural language.
\Cref{fig:interaction} (b) shows that participants relied more on built-in property widgets than on LLM-generated widgets for modifications, partly because the tasks did not require extensive open-ended styling and participants were often satisfied with the initial LLM output without further widget adjustment.
That said, some participants preferred delegating even widget-accessible tasks to NLI when the manual alternative involved coordinating multiple controls.
P7 commented that \textit{``The background as a good thing to leave to the LLM. Adding it manually means adjusting many elements, so it was better to just let the LLM handle it.''}

\subsubsection{Deictic Interaction Builds Modification Confidence}
9 of 12 participants identified deictic interaction as the most advantageous feature of \sys{}.
The core benefit was that pointing to an element before issuing a command reduced the ambiguity inherent in purely verbal references.
P3 explained that \textit{``in DynaVis, I had to spell out things like `the dashed line I created earlier,' but here I just click and say `change this.'''}
P10 similarly noted that \textit{``describing one element at a time in natural language was cumbersome, and when I gave commands about the whole chart, the model got confused and sometimes modified unrelated elements.''}
Log data supported these accounts, with the mean prompt length dropping from 52.3 words in DynaVis ($n{=}240$) to 40.7 words in \sys{} ($n{=}144$), a significant reduction ($p{=}0.005$).
This directness also shaped participants' expectations of success.
P3 reported that \textit{``because I could point to the target, my confidence that the system would actually change it was much higher.''}

Conversely, several participants described anxiety when relying solely on natural language in DynaVis.
P12 noted that \textit{``I had to describe what to change from scratch, and I kept worrying whether the system would understand me.''}
P7 identified deictic reference as the most valuable interaction in \sys{}. 
These observations suggest that deictic interaction not only reduces referential overhead but also lowers the perceived risk of issuing modification commands, encouraging participants to attempt refinements they might otherwise avoid.

\subsubsection{Provenance Provides Spatial Context and Safety}
7 of 12 participants reported the provenance history as beneficial, with 2 identifying it as the most valuable feature overall.

\noindent\textit{Spatial context for revisitation.}
Participants valued that the provenance view spatially organized their modification history, allowing them to maintain context without mentally tracking dependencies.
P3 noted that \textit{``even as the stack grew, the flow was immediately visible, so cognitive load never felt heavy. I could just focus on the current view.''}
P10 similarly observed that \textit{``because modifications are shown as a flow rather than a vertical list, I could follow back and revise without losing context.''}
Participants also found revisiting earlier modifications straightforward: selecting a prior node surfaced the widgets used at that point, so adjustments could be made in place.
P4 remarked that \textit{``there was always a clear point to return to when something went wrong, and finding what I wanted to fix was easy.''}
In contrast, participants found DynaVis's accumulated widgets hard to navigate.
P7 noted that \textit{``in baseline, the history of what I had done was invisible; only the current stacked widgets were shown, so the widgets felt worthless to revisit.''}

\noindent\textit{Psychological safety.}
Beyond practical utility, the provenance view gave participants a sense of security.
P2 stated that \textit{``just seeing the history there was reassuring. It felt reliable.''}
P10 described feeling \textit{``a sense of agency in \sys{}, knowing I could always trace back and revise.''} 
P3 explained that \textit{``the branching structure made history visible and gave me confidence that I could recover from mistakes,''} a feeling absent in DynaVis where widgets accumulated without clear association to prior states.
These observations suggest that provenance not only improves inefficient iterative processes but also encourages bolder design exploration by reducing the perceived cost of experimentation.

\section{Discussion}
\subsection{Visualization as a Graphic Artifact}

Element-level direct selection and manipulation are familiar staples of graphic editors, yet largely absent from visualization authoring tools.
This absence reflects how the field conceives of a chart: as a representation of data, with interactions designed around exploration, querying, and encoding.
In presentation and publication, however, a chart is also consumed as a graphic artifact, where layout, styling, and visual appeal carry communicative weight~\cite{borkin2013memorable, bateman2010usefuljunk}.
Our formative study reflects this duality: practitioners repeatedly left their tools to restyle individual elements, treating the chart as an object to be designed rather than only a view to be queried~\cite{bigelow2014reflections, walny2019datachanges}.

Read through this lens, the techniques we adopt are familiar ones turned to a different end.
Selection and query relaxation, designed to interrogate data and often unable to reach non-data elements such as axes or legends, become a data-aware scope that also targets graphical components; deictic reference, used elsewhere to convey analytical intent~\cite{Chen2025interchat}, instead carries a rendered target into a styling command.
What separates this from mere repurposing is that every edit is written back into the specification: where existing tools force a choice between editing an element as a graphic and keeping it bound to data, \sys{} resolves the gesture into a durable rule on the data-bound specification, so manipulating a chart as a graphic no longer severs it from its data.

This reframing suggests how refinement environments might be designed.
Throughout \sys{}, participants treated refinement as a reversible, low-risk activity rather than destructive, exploring more boldly than their usual workflows allowed.
The broader implication of \sys{} is less a set of techniques than an argument that the refinement stage deserves interactions designed for the artifact view, yet grounded in the data view.

\subsection{Declarative Specification Dependency}
The current implementation operates exclusively on Vega-Lite specifications.
The system's core mechanisms read and write paths within the Vega-Lite JSON structure, so chart types outside Vega-Lite's expressiveness fall beyond the current scope.
While Vega-Lite grammar covers a broad range of chart types and the visual details beyond the marks alone, this expressiveness is still bounded:
\sys{} can refine any chart the grammar expresses but no more, trading expressiveness for refinement that never severs the link between data and visual representation.
This bound is not unique to our approach; any system that maintains data binding through specification-level modification requires a declarative grammar that preserves mappings between data fields and visual encodings~\cite{rahman2025annogram}.
Consequently, refinements that graphic tools support but the grammar cannot express, such as free-form vector annotations or composing separate charts into a single figure with manual layout, remain out of scope.
Other declarative formats such as Vega, ECharts, or Plotly's JSON schema could serve as alternative backends, and an intermediate layer that reconstructs specification-level semantics for imperative graphics libraries could extend the same guarantee to a wider design space. We leave both directions to future work.

\subsection{Data Wrangling and Authoring}
Our system assumes a chart whose data transformation and visual encoding are already established, focusing on post-render refinement.
This scoping decision reflects our research focus on the Design Refinement stage identified in the extended InfoVis Reference Model (\Cref{sec:model}).
However, in practice, refinement rarely occurs in isolation from earlier pipeline stages.
Participants in our formative study described workflows where data wrangling, encoding decisions, and visual refinement interleave, particularly when data updates require re-encoding or when a refinement exposes an issue in the data structure.

Supporting data transformation and initial chart authoring within the same environment would allow users to maintain a single, continuously data-bound workflow from raw data through to presentation-ready output, fully realizing the connected pipeline our extended model calls for.
Since Vega-Lite specifications are expressed in a JSON-based declarative grammar that already includes data transformation operators such as filter, aggregate, and fold, extending the system in this direction is a feasible next step.
The interaction patterns introduced in \sys{}, particularly scope expansion and deictic reference, could also apply to authoring tasks, such as pointing at a data column to define an encoding or expanding a filter scope to a categorical group.

\subsection{Deployment study}
Our user study involved short sessions with predefined tasks that did not fully exercise all interaction techniques.
Several participants noted that they did not use provenance branching because the tasks did not require extensive exploration.
A longer-term deployment in participants' actual work environments would reveal how these features are adopted in practice, particularly for complex, open-ended refinement scenarios where provenance and iterative exploration become more critical.

\section{Conclusion}
In this paper, we identified a missing Design Refinement stage in the traditional InfoVis Reference Model through a formative study with 18 practitioners and a follow-up survey of 35 respondents.
Our findings show that current tools often force practitioners to choose between expressive visual control and data-binding integrity.
We presented \sys{}, a chart refinement system that supports expressive design refinement while preserving data-binding integrity.
Based on the formative study findings, \sys{} proposes complementary interaction techniques, introducing data-aware scope expansion, deictic interaction for NLI and dynamic widgets, all integrated with a provenance history.
A user study confirmed significant improvements in refinement completeness, expressiveness, and controllability over a validated baseline, demonstrating that closing the gap between rendered view and presentation-ready output is both necessary and achievable within a single data-bound environment.

\acknowledgments{This work was supported by the National Research Foundation of Korea (NRF) grant funded by the Korean government (MSIT) (No.NRF-2023R1A2C2005209), the Institute of Information \& communications Technology Planning \& Evaluation (IITP) grant funded by the Korean government (MSIT) [No.RS-2021-II211343, Artificial Intelligence Graduate School Program (Seoul National University)], the AI Seoul Tech Research Support Program of the Seoul Future Foundation, and the SNU-Global Excellence Research Center establishment project. The ICT at Seoul National University provided research facilities for this study.}

\bibliographystyle{abbrv-doi-hyperref}

\bibliography{ref}

\end{document}